\documentclass[aps,prl,twocolumn,superscriptaddress,amsmath,amssymb,floatfix]{revtex4-2}

\usepackage{graphicx}
\usepackage{bm}
\usepackage{xcolor}
\usepackage[colorlinks=true,linkcolor=blue,citecolor=blue,urlcolor=blue]{hyperref}
\usepackage{siunitx}

\newcommand{\Kp}{K'}
\newcommand{\vF}{v_{\mathrm F}}

\begin{document}

\title{Valley-resolved high harmonics probe symmetry breaking beyond the Dirac cone}

\author{Pablo Pais}
\affiliation{Institute of Particle and Nuclear Physics, Faculty of Mathematics and Physics, Charles University, V Hole\v{s}ovi\v{c}k\'{a}ch 2, 18000 Prague 8, Czech Republic}
\affiliation{Instituto de Ciencias F\'isicas y Matem\'aticas, Universidad Austral de Chile, Casilla 567, 5090000 Valdivia, Chile}

\author{Alfredo Iorio}
\affiliation{Istituto Nazionale di Fisica Nucleare, Bologna branch, Via Berti Pichat 6/2, 40127 Bologna, Italy}
\affiliation{Institute of Particle and Nuclear Physics, Faculty of Mathematics and Physics, Charles University, V Hole\v{s}ovi\v{c}k\'{a}ch 2, 18000 Prague 8, Czech Republic}

\author{Marcelo F. Ciappina}
\email{marcelo.ciappina@gtiit.edu.cn}
\affiliation{Department of Physics, Guangdong Technion-Israel Institute of Technology, Shantou 515063, China}
\affiliation{Guangdong Provincial Key Laboratory of Materials and Technologies for Energy Conversion, Guangdong Technion-Israel Institute of Technology, Shantou 515063, China}

\date{\today}

\begin{abstract}
Low-energy electrons in graphene obey an emergent, rotationally invariant Dirac equation, while the honeycomb lattice enters only through higher-order, symmetry-breaking corrections, the leading one being trigonal warping. 
This provides a unique solid-state arena to address a cornerstone challenge in quantum gravity: how to detect ultraviolet symmetry-breaking granular corrections, suppressed by a fundamental scale, atop an emergent infrared relativistic continuum.
We show that intense counter-rotating $\omega$–$2\omega$ fields and valley-resolved high-harmonic generation (HHG) provide such a probe. Within a family of models that interpolates continuously between the isotropic Dirac cone and nearest-neighbour graphene at fixed Fermi velocity, the valley contrast of the harmonic yield vanishes identically in the symmetric theory, grows linearly with the warping coupling, and its valley-even counterpart appears only at second order, as expected from effective-field-theory power counting. 
The stronger the laser, the farther it drives electrons from the Dirac points, so the laser intensity plays the role of the collision energy. For the fourth harmonic, the leading warping correction alone accounts for the valley contrast of real graphene up to fundamental intensities of about $10^{11}$~W/cm$^2$, below the damage threshold of monolayer graphene, whereas the fifth harmonic already reveals higher-order lattice corrections at intensities around $10^{10}$~W/cm$^2$. The two harmonics, which rotate in opposite directions, are emitted by opposite valleys. Valley-resolved harmonics are far more sensitive than the residual valley populations, making them a practical detector of lattice-scale physics hidden beneath an emergent Dirac theory.
\end{abstract}

\maketitle

\paragraph{Introduction.}

{  Symmetries that are exact at low energies but broken at a high scale constitute a central and recurring theme in quantum gravity (QG)~\cite{AmelinoCamelia1999Nature}. While special relativity is widely verified in the infrared (IR) limit, modern QG scenarios suggest that continuous Lorentz covariance might only be an emergent property of a smoother spacetime continuum~\cite{AmelinoCamelia1998Nature}. At the ultraviolet (UV) Planck scale, the discrete, granular nature of spacetime is predicted to emerge, naturally leading to a breakdown of continuous spacetime symmetries and motivating extensions of the Standard Model~\cite{Colladay1998, Kostelecky2011, Myers2003}. This Planck-scale granularity can be encoded via modified Heisenberg algebras that introduce a fundamental minimal length through Generalized Uncertainty Principles (GUPs)~\cite{Kempf:1994su}. Recently, major theoretical and phenomenological efforts have focused on tracking and constraining these GUP-induced effects in macroscopic systems~\cite{Petruzziello:2020wkd, Bosso:2023aht} and highly energetic cosmic or collider frameworks~\cite{Mureika:2026khw}. From an effective field theory (EFT) perspective, the major challenge lies in designing observables where the low-energy symmetric theory predicts an exact null result, allowing any small, high-energy, symmetry-violating signature to be measured against a clean background.}

{  Condensed-matter systems, where pseudo-relativistic structures emerge from a discrete lattice, offer a fully controlled, high-precision arena for exploring these quantum-gravitational ideas~\cite{Barcelo2011}. Dirac materials, with graphene monolayer at the forefront~\cite{Semenoff1984, Novoselov2005, CastroNeto2009}, are the cleanest examples: near the $K$ and $K'$ corners of the Brillouin zone, conducting ($\pi$) electrons behave as emergent massless Dirac fermions with velocity $v_F \approx c/300$. While early works successfully utilized this linear regime to simulate astrophysical setups, such as curved spacetimes and Hawking-Unruh phenomena~\cite{Iorio2014, iorio2018anti}, a deeper analogy emerges when exploring the system beyond the linear regime. As energy increases, the graphene lattice spacing $l_{\text{graphene}} \sim \SI{e-10}{\meter}$ plays the role of a \textit{macroscopic Planck length}, where the underlying discrete lattice acts as the granular quantum space. A generalized, non-linear Dirac structure survives at higher energies~\cite{Iorio2017GUP, Iorio2019proc}, and it can be described using a set of generalized phase-space variables, organized into a multi-layered energy hierarchy~\cite{threelayers}.  In this framework, the high-energy corrections, such as the particle-hole asymmetry and the anisotropic trigonal warping~\cite{Ando1998, McCann2006}, can be mapped onto GUP-modified commutation relations~\cite{threelayers} and non-commutative spaces~\cite{Iorio2023noncommutativity}. Consequently, driving electrons far from the Dirac cones with intense laser fields turns the platform into a tabletop particle accelerator for probing the UV breaking of emergent continuous symmetries.}

Strong laser fields, the workhorse of attosecond science~\cite{Corkum1993,Lewenstein1994,KrauszIvanov2009,Ciappina2017}, can drive electrons far from the Dirac points within a
fraction of an optical cycle and induce nonequilibrium transport~\cite{Higuchi2017,Kelardeh2015,McIver2020}, and the emitted
high harmonics~\cite{Ghimire2011,Schubert2014,Vampa2014,Luu2015,Vampa2015Nature,Hohenleutner2015,GhimireReis2019} encode the
band structure and Berry curvature they explore~\cite{TancogneDejean2017,LuuWorner2018}. In graphene, the interplay of intra- and interband dynamics produces a rich nonperturbative response~\cite{AlNaib2014,Taucer2017,Chizhova2017,Yoshikawa2017,Baudisch2018,MrudulDixit2021PRB}. {  This nonperturbative response is particularly relevant, given that $\pi$ electrons propagating through the underlying lattice with dislocations can form effective time loops, whose geometric holonomies can be probed and distinguished via laser-driven high-harmonic generation~\cite{Ciappina2019prd, Iorio2020tuc}.} In particular, counter-rotating bicircular
$\omega$--$2\omega$ fields~\cite{Eichmann1995,Milosevic2000,Fleischer2014}, whose trefoil Lissajous figure matches the threefold
symmetry of the lattice~\cite{Neufeld2019,Saito2017}, induce
valley polarization in pristine, gapless graphene and imprint it on the
harmonic spectrum~\cite{Mrudul2021Optica,Mrudul2021JPB}. Lightwave
valleytronics has been demonstrated in a gapped monolayer~\cite{Langer2018},
and strong fields have been proposed to control valley and topological
properties of gapped and topological materials on sub-cycle
timescales~\cite{Silva2019,JimenezGalan2020NatPhoton}. The effect has been attributed
qualitatively to the anisotropic valley landscape away from the Dirac
points~\cite{Mrudul2021Optica}.

Here we make this attribution quantitative and recast it in the language of
effective field theory (EFT). We introduce a one-parameter family of
Hamiltonians that switches the lattice corrections on continuously at fixed
Fermi velocity, show that the isotropic Dirac theory is a symmetry-protected
null test, and demonstrate that the valley-resolved harmonic yield obeys the
power counting expected for a dimension-raising, rotation-breaking operator.
The vector potential of the driving field sets the probed momentum, so that
the field strength plays the role of the collision energy, and we identify
where the expansion breaks down.

\begin{figure*}[tb]
\includegraphics[width=0.9\textwidth]{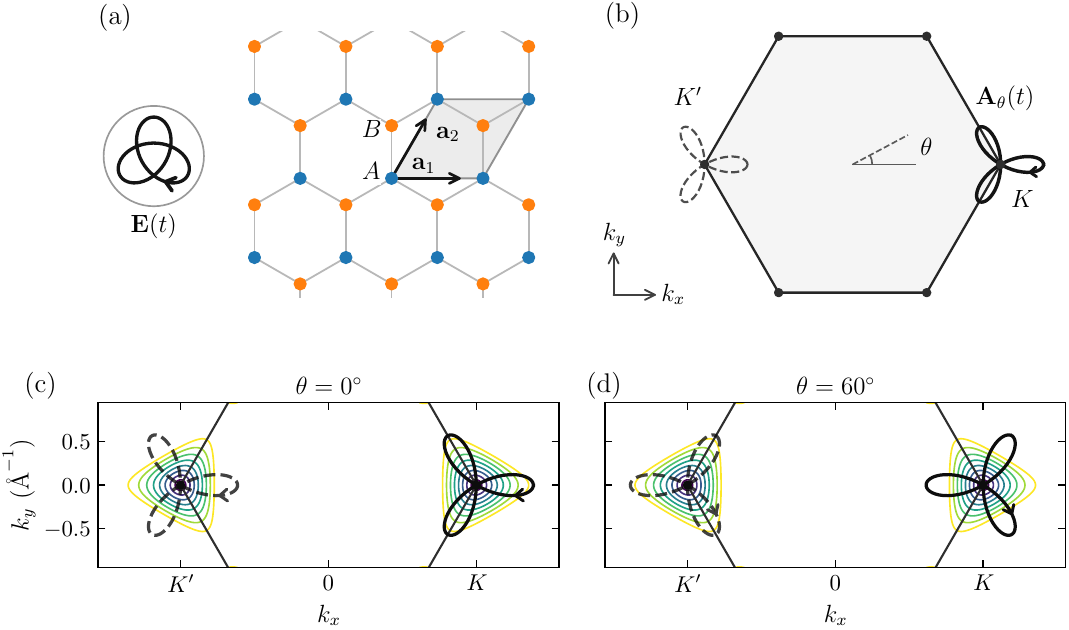}
\caption{(a) Honeycomb lattice and counter-rotating bicircular field
$\bm E(t)$. (b) Brillouin zone and the vector-potential trefoil
$\bm A_\theta(t)$ drawn at $K$ (solid) and $\Kp$ (dashed); $\theta$ is the
orientation of the field relative to the lattice. (c), (d) Constant-energy contours of the
tight-binding bands around $K'$ and $K$, showing trigonal warping of opposite
orientation, with the trefoil for $\theta=0^\circ$ and $60^\circ$.}
\label{fig:setup}
\end{figure*}

\paragraph{Model.}
We describe graphene by the nearest-neighbour tight-binding Hamiltonian
$H(\bm k)=-t\,[f(\bm k)\,\sigma_+ + \mathrm{h.c.}]$, with
$f(\bm k)=\sum_{j=1}^{3}e^{i\bm k\cdot\bm\delta_j}$, hopping $t=2.7$~eV and
carbon--carbon distance $a_{cc}=1.42$~\AA~\cite{Reich2002,CastroNeto2009}. To
separate the emergent Dirac theory from its lattice completion we introduce
the family
\begin{equation}
H_\lambda(\bm k) = -\frac{t}{\lambda}\Big[f\big(\bm K_\tau+\lambda\bm q\big)\sigma_+
+\mathrm{h.c.}\Big],\qquad \bm k=\bm K_\tau+\bm q ,
\label{eq:Hlambda}
\end{equation}
obtained by scaling $a_{cc}\to\lambda a_{cc}$ and $t\to t/\lambda$. The Fermi
velocity $\hbar\vF=\tfrac32 t a_{cc}$ is independent of $\lambda$. Expanding
around valley $\tau=\pm1$,
\begin{equation}
H_\lambda = \hbar\vF\,(\tau q_x\sigma_x+q_y\sigma_y)
+\lambda\,\frac{\hbar\vF a_{cc}}{4}\,\mathcal O^{(2)}_\tau(\bm q)
+\mathcal O(\lambda^2 q^3),
\label{eq:eft}
\end{equation}
where $\mathcal O^{(2)}_\tau$ is the quadratic trigonal-warping
term~\cite{Ando1998,McCann2006}, which carries opposite sign in the two
valleys. The dispersion becomes
$E=\pm\hbar\vF q\,[1-\tau\lambda (a_{cc}q/4)\cos 3\vartheta_{\bm q}+\dots]$,
with $\vartheta_{\bm q}$ measured from a lattice axis, so that the
constant-energy contours of $K$ and $\Kp$ are triangles of opposite
orientation [Fig.~\ref{fig:setup}(c,d)]. Equation~\eqref{eq:eft}
is the analogue of a Lorentz-invariant theory supplemented by a higher-dimension operator with coupling $\lambda/\Lambda$,
$\Lambda=4/a_{cc}\approx2.8$~\AA$^{-1}$.
In the language of quantum gravity, the scaling parameter $\lambda$ acts as an effective regulator, with $\lambda\to0$ corresponding to the smooth, isotropic IR ``world'', whereas activating the $\mathcal{O}(\lambda)$ trigonal-warping correction marks the entrance into the deformed ``superworld'' and ``hyperworld'' layers (~\cite{threelayers}).

The driving field is a counter-rotating bicircular pulse with vector
potential
$\bm A(t)=\frac{A_0}{\sqrt2}g(t)\big[(\cos(\omega t+\phi),\sin(\omega t+\phi))
+\frac R2(\cos2\omega t,-\sin2\omega t)\big]$, with $\lambda_L=6~\mu$m,
$R=2$, a $\sin^2$ envelope $g(t)$ of total duration 145~fs, and field-strength
parameter $F$ between 5 and 15~MV/cm ($A_0=eF/\hbar\omega=0.73$~\AA$^{-1}$
at 15~MV/cm). Its electric field and vector potential trace three-leaved
trefoils [Fig.~\ref{fig:setup}(a,b)], whose orientation relative to the
warped valleys is set by $\phi$ or, equivalently, by the crystal angle
$\theta$ [Fig.~\ref{fig:setup}(c,d)]. We solve the semiconductor Bloch equations in the
Houston picture, $\bm k(t)=\bm q+\bm A(t)$, with interband
dephasing $T_2=10$~fs and no population relaxation~\cite{Golde2008,Floss2018,Mrudul2021Optica}.
The current $\bm J_\tau(t)=\sum_{\bm q}\mathrm{Tr}[\rho\,\partial_{\bm k}H_\lambda]$
is accumulated separately for each valley. Over a single valley region this
sum also contains the current that the filled valence band carries when it
is rigidly displaced by $\bm A(t)$. That term vanishes for the complete
Brillouin zone but not for a valley region, where it is fixed by the
arbitrary position of the boundary. We subtract it, so that $Y_N^{K}$ and
$Y_N^{\Kp}$ (integrated over $N\pm0.25$) are the yields of the carriers and
interband coherences created in each valley (Supplemental Material). For the
interpolation in $\lambda$, both valleys are sampled on identical discs
$|\bm q|\le1.3$~\AA$^{-1}$, reduced to half the $K$--$\Kp$ distance of the
scaled lattice, $0.85/\lambda$~\AA$^{-1}$, when the discs would otherwise
overlap ($\lambda>0.65$); with this definition the H4 yields change by only a
few percent when the disc radius is varied. Full-Brillouin-zone calculations
for $\lambda=1$ use valley sectors with up to $85^2$ points, and the valley
contrasts change by less than $0.004$ between $65^2$ and $85^2$ points
(Supplemental Material).

\begin{figure*}[tb]
\includegraphics[width=0.85\textwidth]{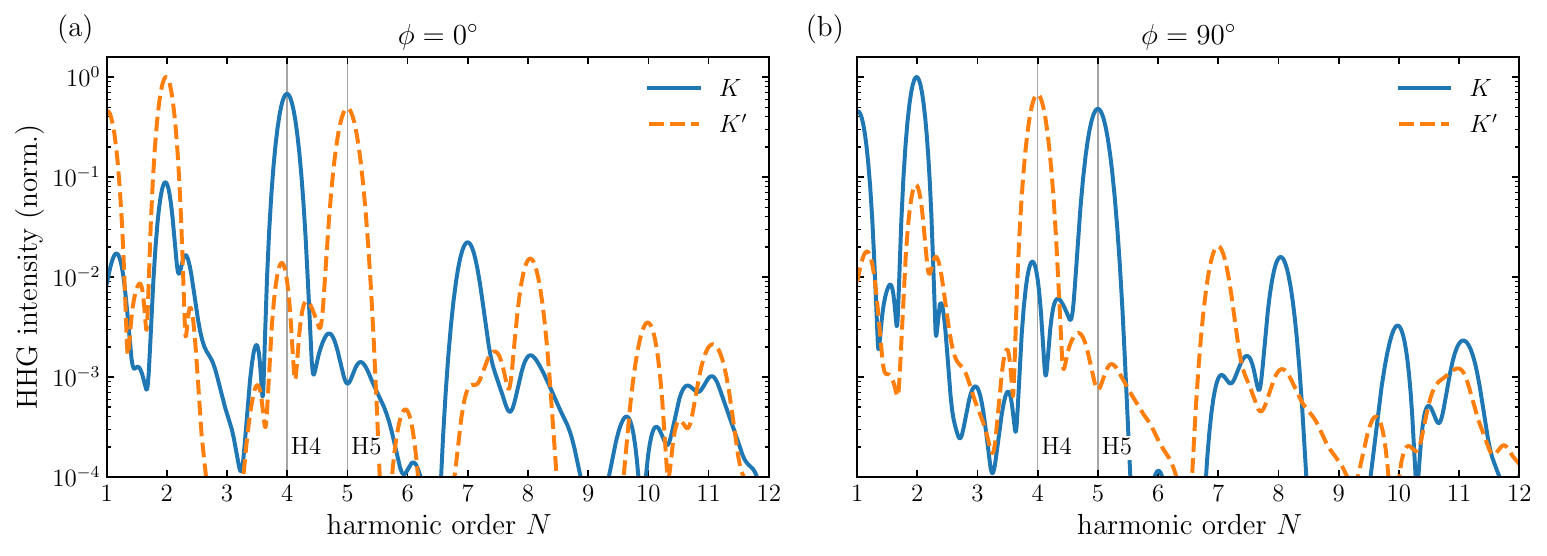}
\caption{Valley-resolved harmonic spectra of graphene ($\lambda=1$,
$F=15$~MV/cm, full Brillouin zone) for two-colour phase (a) $\phi=0^\circ$
and (b) $\phi=90^\circ$. H4 and H5 switch between the $K$ (solid) and $\Kp$
(dashed) valleys as $\phi$ advances by $90^\circ$, equivalent to a
$60^\circ$ rotation of the field. Spectra are computed from the valley
currents, after removing the filled-band current of each valley, with a Hann
window and 16-fold zero padding; no smoothing is applied.}
\label{fig:spectra}
\end{figure*}

\paragraph{Valley-resolved harmonics in graphene.}
Figure~\ref{fig:spectra} shows the valley-resolved spectra of full graphene.
The trefoil field generates the $3n\pm1$ harmonics allowed by its dynamical
symmetry~\cite{Neufeld2019,Saito2017}, and the two valleys radiate very
differently. At $\phi=0$, $F=15$~MV/cm, the $K$ valley dominates H4 by a
factor of $\approx60$, whereas the $\Kp$ valley dominates H5 by a factor of
$\approx240$; at $\phi=90^\circ$ the roles are exchanged. We quantify this by the
harmonic valley contrast $P_N=(Y_N^{\Kp}-Y_N^{K})/(Y_N^{\Kp}+Y_N^{K})$, which
reaches $P_4=-0.97$ and $P_5=+0.99$. The two harmonics thus come from opposite
valleys. Since H4 co-rotates with the fundamental and H5 with the second
harmonic, each valley emits predominantly one helicity; we find this
valley--helicity locking at every phase, field strength and dephasing time
studied, whenever the contrast is appreciable. For comparison, the residual
conduction-band valley asymmetry $\eta=(n_c^{\Kp}-n_c^K)/\bar n_c$ under the
same conditions is $-0.37$, and at $F=5$~MV/cm it drops to $1.6\%$ while
$|P_4|$ remains $0.98$. Valley-resolved harmonics are therefore a far more
sensitive detector than residual populations, whose value moreover depends
on how the valley regions are delimited (Supplemental Material).

\begin{figure}[t]
\includegraphics[width=\columnwidth]{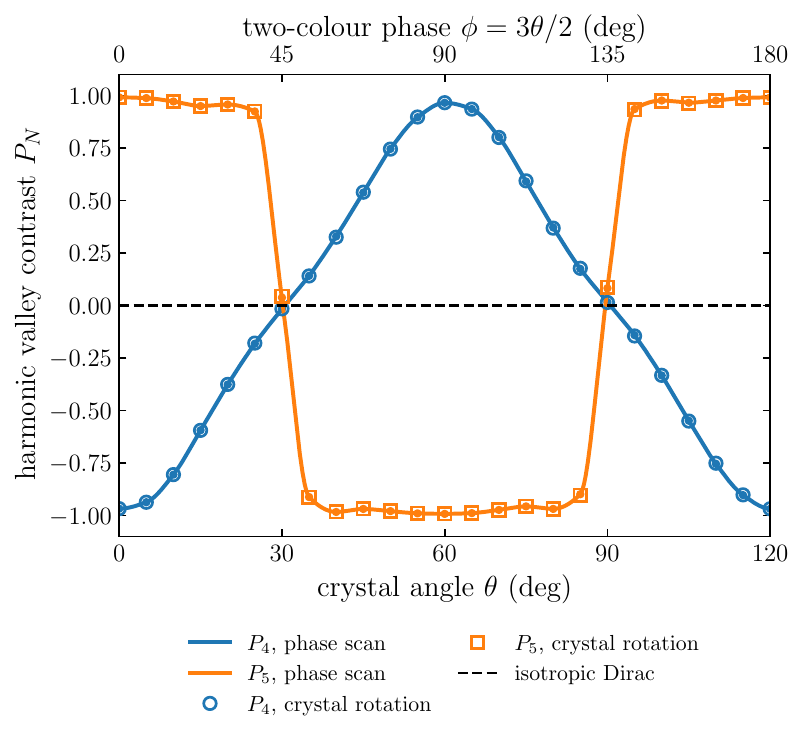}
\caption{Harmonic valley contrast of graphene ($F=15$~MV/cm, full Brillouin
zone) versus crystal angle $\theta$ (open symbols) and versus two-colour phase
$\phi$ plotted at $\theta=2\phi/3$ (dots, joined by a shape-preserving
interpolant). The two scans coincide, as
required by Eq.~\eqref{eq:rotphase}. The isotropic Dirac model gives
$P_N\equiv0$ (dashed) for every $\theta$ and $\phi$.}
\label{fig:c3}
\end{figure}

\paragraph{The Dirac theory as a symmetry-protected null test.}
Two symmetry statements organize the problem. First, a rotation of the
bicircular field by an angle $\theta$ is equivalent to a time shift
$\theta/2\omega$ together with a change of the two-colour phase,
\begin{equation}
\phi\to\phi+\tfrac32\theta ,
\label{eq:rotphase}
\end{equation}
up to the slowly varying envelope. A crystal
rotation and a phase scan therefore probe the same physics. Our numerics
confirm this to within $0.01$: for instance, $P_4(\theta=10^\circ)=-0.805$ and
$P_4(\phi=15^\circ)=-0.805$, $P_4(\theta=20^\circ)=-0.375$ and
$P_4(\phi=30^\circ)=-0.376$, $P_4(\theta=60^\circ)=+0.966$ and
$P_4(\phi=90^\circ)=+0.966$ (Fig.~\ref{fig:c3}). Second, the $\lambda=0$
theory is invariant under continuous rotations, and its two valleys are
mirror images. Combining these, every observable of the isotropic Dirac model
is independent of $\theta$ and $\phi$, and every valley-odd observable
vanishes identically. Our Dirac calculations reproduce this to machine
precision ($|P_N|<10^{-15}$ for all $\theta$).

The vanishing of $P_N$ in the Dirac model is therefore not a numerical finding
but a selection rule, exactly as the vanishing of a Lorentz-violating
coefficient in the Standard Model~\cite{Colladay1998,Kostelecky2011}. Its
value lies in the converse: any nonzero valley contrast, and any dependence
of a harmonic yield on $\phi$ or on the crystal orientation, is a direct
measurement of rotation-breaking corrections to the Dirac theory. The
threefold modulation of $P_N(\theta)$ in Fig.~\ref{fig:c3} is thus an
interferometric readout of the lattice. What it does not yet tell us is
\emph{which} correction is responsible and how its effect scales; we
address this next.

\paragraph{Power counting of the warping operator.}
The family~\eqref{eq:Hlambda} obeys an exact identity,
$H^{\Kp}_\lambda=\sigma_yH^{K}_{-\lambda}\sigma_y$, which follows from
$f(-\bm k)=f^*(\bm k)$ (Supplemental Material). The valley $\Kp$ at coupling
$\lambda$ is thus unitarily equivalent to the valley $K$ at coupling $-\lambda$,
so that $Y_N^{\Kp}(\lambda)=Y_N^{K}(-\lambda)$ for any driving field. At
$\lambda=0$ this is the null result above; for $\lambda\neq0$ it fixes the
parity of every valley observable in the coupling. Physically, the
valley-resolved emission amplitude at harmonic $N$ is
$a_N^{(0)}+\tau\lambda a_N^{(1)}+\mathcal O(\lambda^2)$,
where $a_N^{(0)}$ is the Dirac amplitude, identical in both valleys, and the
sign $\tau$ reflects the opposite warping orientation at $K$ and $\Kp$. The
yields then decompose into a valley-odd part linear in the coupling and a
valley-even part quadratic in it,
\begin{align}
D_N &\equiv \frac{Y_N^{\Kp}-Y_N^{K}}{2Y_N^{(0)}} = c_N\,\lambda+\mathcal O(\lambda^3),
\label{eq:D}\\
S_N &\equiv \frac{Y_N^{\Kp}+Y_N^{K}}{2Y_N^{(0)}} = 1+d_N\,\lambda^2+\mathcal O(\lambda^4),
\label{eq:S}
\end{align}
with $Y_N^{(0)}$ the Dirac yield of one valley. This is the familiar EFT
pattern: interference with the leading-order amplitude is linear in the new
coupling, while the new amplitude squared enters at second order.

\begin{figure*}[tb!]
\includegraphics[width=\textwidth]{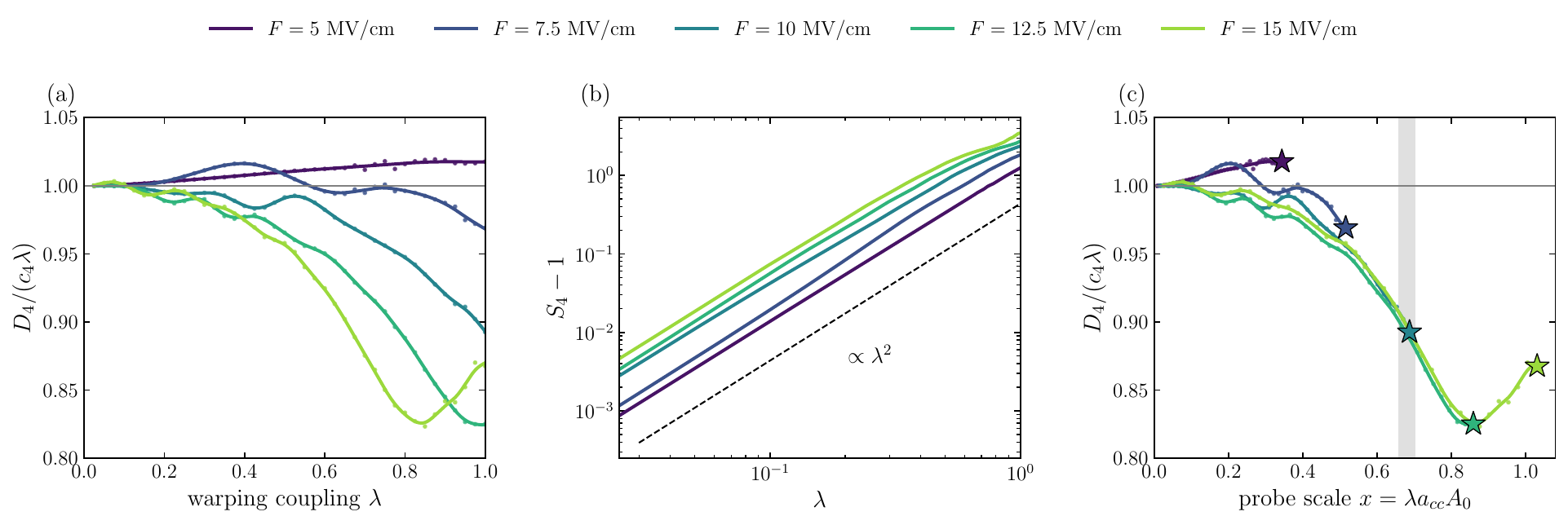}
\caption{Interpolation from the isotropic Dirac cone ($\lambda=0$) to
graphene ($\lambda=1$) at fixed Fermi velocity for H4, $\phi=0$, and five
field strengths. (a) Valley-odd yield normalized to its first-order
form, $D_4/(c_4\lambda)$; unity marks first order in the warping coupling.
(b) Valley-even shift $S_4-1$; the dashed line is $\propto\lambda^2$.
(c) $D_4/(c_4\lambda)$ versus the probe scale $x=\lambda a_{cc}A_0$; the grey
band marks the breakdown of the expansion and the stars mark real graphene
($\lambda=1$). Dots are the computed couplings ($\Delta\lambda=0.025$); lines
are smoothing splines (maximum deviation $0.006$). Refining the momentum grid
and time step, or enlarging the disc, changes $D_4/(c_4\lambda)$ by at most
$3$--$4\%$; the small oscillations of 1--2\% are within this numerical
uncertainty.}
\label{fig:scaling}
\end{figure*}

Figure~\ref{fig:scaling} tests Eqs.~\eqref{eq:D} and~\eqref{eq:S} by scanning
$\lambda$ from the Dirac limit to graphene in steps of $0.025$, for five field
strengths between 5 and 15~MV/cm. Normalizing the valley-odd yield by its
initial slope, $D_4/(c_4\lambda)$ stays at unity over a broad range of
couplings for every field [Fig.~\ref{fig:scaling}(a)]: the valley contrast is
first order in the warping operator. The valley-even shift follows
$S_4-1\propto\lambda^2$ over more than a decade [Fig.~\ref{fig:scaling}(b)];
$(S_4-1)/\lambda^2$ is constant to within $6\%$ for $\lambda\le0.1$ at all
fields. The coefficients $c_N$ and $d_N$ depend on harmonic order and field
strength (for H4, $c_4$ grows in magnitude from $-2.2$ at 5~MV/cm to $-5.0$ at
15~MV/cm), and $c_N$ reverses sign when $\phi$ is advanced by $90^\circ$, which
exchanges the valleys.

\paragraph{The field as collision energy.}
The driving field displaces the crystal momentum by $\bm A(t)$, so the
largest momentum probed is $\sim A_0$ and the natural expansion parameter of
Eq.~\eqref{eq:eft} is
\begin{equation}
x=\lambda\,a_{cc}A_0=\frac{4\lambda A_0}{\Lambda},
\end{equation}
the analogue of $E/M$. In Fig.~\ref{fig:scaling}(c) we plot the normalized
first-order coefficient against $x$. The curves for all five fields collapse
onto a single function of $x$, the hallmark of a single expansion parameter.
The first-order description holds to within $5\%$ up to $x=0.5$ and breaks down,
with deviations exceeding $10\%$, at $x\simeq0.7$ (grey band). Real graphene
($\lambda=1$, stars) corresponds to $x=0.069\,F$[MV/cm] at $6~\mu$m, i.e.\
$x=0.34$, $0.69$ and $1.03$ for 5, 10 and 15~MV/cm. Up to about 10~MV/cm the H4
valley contrast of real graphene is therefore accounted for, to within $10\%$,
by the leading warping operator alone, while stronger fields begin to resolve
the full lattice, the analogue of the ultraviolet completion. The H5 contrast
collapses onto a universal curve as well, but departs from first order already
at $x\approx0.2$ and grows to twice the first-order value at $x\approx1$
(Supplemental Material): the two harmonics thus probe different orders of the
expansion. For $\lambda=1$ the valley contrasts obtained on the Dirac-centred
discs agree with the full-zone values (for example $P_4=-0.975$ versus
$-0.967$ at 15~MV/cm).

\paragraph{Experimental access.}
The coupling $\lambda$ is a theoretical device, but its role can be
mirrored experimentally in two ways. First, the field strength tunes $x$
continuously, so the transition between the linear and lattice-dominated
regimes is accessible in a single sample by scanning the mid-infrared
intensity. Second, valley-resolved yields need not be measured directly:
because a $90^\circ$ step in $\phi$ exchanges the valleys, the total yield
$Y_N^{K}+Y_N^{\Kp}$ is modulated with a $90^\circ$ period in $\phi$ in
graphene, whereas in the isotropic theory it is independent of $\phi$. In our
full-zone calculations at 15~MV/cm the total H5 yield varies by a factor of
$\approx5$ over a phase scan, a signal that only requires control of the
$\omega$--$2\omega$ delay. Its helicity provides a further handle, since
the $3n+1$ and $3n+2$ harmonics co-rotate with the $\omega$ and $2\omega$
components, respectively~\cite{Fleischer2014,Saito2017}. Related
nonlinear-optical signatures of valley polarization, such as
valley-induced second-harmonic generation~\cite{Golub2014}, offer
complementary readouts. Materials with different warping strength, or
graphene under strain~\cite{Pereira2009}, would provide additional ``couplings''.
In a finite pulse, a rotation and a phase shift differ by a shift of the
carrier relative to the envelope, so even the isotropic model shows a weak
residual $\phi$ dependence of the total yield (a few percent in our
calculations), which sets the floor against which the lattice-induced
modulation should be compared.

\paragraph{Conclusion.}
Graphene offers a rare setting in which an emergent relativistic theory and
its symmetry-breaking ultraviolet completion are both known exactly. We have
shown that bicircular strong-field driving turns this into an experiment:
the isotropic Dirac theory predicts an exact null valley contrast, the
trigonal-warping operator lifts it linearly in its coupling and with the
valley-even response at second order, and the laser field strength scans
the probe scale across the breakdown of the effective description. The
fourth harmonic measures the leading coupling, the fifth reveals higher-order
lattice terms, and the two are emitted by opposite valleys.
Valley-resolved high harmonics are thus a sensitive detector of
lattice-scale physics hidden beneath emergent Dirac dynamics, and a natural
laboratory analogue of searches for higher-dimension, symmetry-breaking
operators in high-energy physics~\cite{Myers2003,Kostelecky2011}. The same
strategy applies to other Dirac and Weyl materials~\cite{Wehling2014,Armitage2018},
where tilt, anisotropy, or mass terms play the role of different
symmetry-breaking couplings.

{  Our results provide an operational verification of the multi-layered generalized Heisenberg algebras framework, analytically developed for graphene's lattice completion~\cite{Iorio2017GUP,threelayers,Iorio2023noncommutativity}. The exact vanishing of the valley contrast \(P_N\) in the pristine Dirac limit ($\lambda=0$) represents a symmetry-protected null test that physically materializes the standard, undeformed Heisenberg algebra. Conversely, the strictly linear scaling of the fourth harmonic (\(D_4 \propto \lambda\)) acts as a direct macroscopic readout of the first-order GUP corrections, mapping the algebraic transition from the canonical laboratory variables to the intermediate ``supervariables'' layer~\cite{Iorio2017GUP,threelayers}.
Furthermore, the early departure from linearity observed in the fifth harmonic ($H_5$) at $x \approx 0.2$ serves as a clear signature of the non-linear, higher-order lattice operators that govern the deep ``hyperworld'' regime~\cite{threelayers}. Ultimately, isolating these specific harmonic signatures provides the necessary toolset for optically probing more intricate geometric holonomies, such as the effective time loops formed by $\pi$ electrons in dislocated graphene lattices~\cite{Ciappina2019prd}.}

\begin{acknowledgments}
M.~F.~C. acknowledges support by the Quantum Science and Technology-National Science and Technology Major Project (Grant No. 2025ZD0301000),  the National Key Research and Development Program of China (Grant No. 2023YFA1407100), the Guangdong Province Science, Technology Major Project (Future functional materials under extreme conditions - 2021B0301030005), the National Natural Science Foundation of China (Grant No. 12574092) and from the Knut and Alice Wallenberg Foundation through the Wallenberg Centre for Quantum Technology (WACQT).
{  A.~I. and P.~P. gladly acknowledge the financial support of Charles University Grant UNCE 24/SCI/016.}
\end{acknowledgments}

\bibliography{refs}

\end{document}